\documentclass{iopart}

\usepackage{pslatex}
\usepackage{iopams}
\usepackage[dvips]{graphicx}

\begin{document}
\title[Precise calibration using photon radiation pressure]{Precise calibration of LIGO test mass actuators using photon radiation pressure}
\author{E~Goetz$^1$, P~Kalmus$^{2}$\footnote{Current affiliation: California Institute of Technology, Pasadena, CA 91125, USA}, S~Erickson$^3$\footnote{Current affiliation: Smith College, Northampton, MA 01063, USA}, R~L~Savage~Jr$^3$, G~Gonzalez$^4$, K~Kawabe$^3$, M~Landry$^3$, S~Marka$^2$, B~O'Reilly$^5$, K~Riles$^1$, D~Sigg$^3$ and P~Willems$^6$}
\address{$^1$ University of Michigan, Ann Arbor, MI 48109, USA} 
\address{$^2$ Columbia University, New York, NY 10027, USA}
\address{$^3$ LIGO Hanford Observatory, Richland, WA 99352, USA}
\address{$^4$ Louisiana State University, Baton Rouge, LA 70803, USA}
\address{$^5$ LIGO Livingston Observatory, Livingston, LA 70754, USA}
\address{$^6$ California Institute of Technology, Pasadena, CA 91125, USA}

\eads{\mailto{egoetz@umich.edu}, \mailto{savage\_r@ligo-wa.caltech.edu}}

\begin{abstract}
Precise calibration of kilometer-scale interferometric gravitational wave detectors is crucial for source localization and waveform reconstruction. A technique that uses the radiation pressure of a power-modulated auxiliary laser to induce calibrated displacements of one of the $\sim$10 kg arm cavity mirrors, a so-called {\it photon calibrator}, has been demonstrated previously and has recently been implemented on the LIGO detectors. In this article, we discuss the inherent precision and accuracy of the LIGO photon calibrators and several improvements that have been developed to reduce the estimated voice coil actuator calibration uncertainties to less than 2 percent ($1\sigma$). These improvements include accounting for rotation-induced apparent length variations caused by interferometer and photon calibrator beam centering offsets, absolute laser power measurement using temperature-controlled InGaAs photodetectors mounted on integrating spheres and calibrated by NIST, minimizing errors induced by localized elastic deformation of the mirror surface by using a two-beam configuration with the photon calibrator beams symmetrically displaced about the center of the optic, and simultaneously actuating the test mass with voice coil actuators and the photon calibrator to minimize fluctuations caused by the changing interferometer response. The photon calibrator is able to operate in the most sensitive interferometer configuration, and is expected to become a primary calibration method for future gravitational wave searches.
\end{abstract}


\section{Introduction}
The Laser Interferometer Gravitational wave Observatory (LIGO) detectors are power-recycled Michelson interferometers with Fabry-Perot arm cavities.~\cite{LIGOdetectors} These detectors can sense differential-length variations with amplitude spectral densities on the order of $10^{-19}$~m$/\sqrt{\textrm{Hz}}$ near 150 Hz.~\cite{IFOsensitivity} Passing gravitational waves that cause space to stretch and compress along the arm cavities would be sensed by the interferometer as differential-length changes of the arm cavities. Feedback control loops are used to maintain the nominal separation of the interferometer mirrors required for gravitational wave detection. In particular, the differential-arm length (DARM) control loop uses magnets glued to the back surfaces of the end mirrors ({\it end test masses} or ETMs), surrounded by coils of wire, to control the positions of the suspended mirrors without mechanically contacting the optics. These magnet-coil pairs are known as {\it voice coil actuators}. This article describes how we use radiation pressure actuators, photon calibrators, to determine the actuation function for these voice coil actuators and discusses the accuracy and precision achieved by the LIGO photon calibrator systems.

Photon calibrators have been implemented at the Glasgow 10-meter prototype detector~\cite{GlasgowPcal} and the GEO600 detector~\cite{GEOPcal}, and they have been under development at LIGO and VIRGO for a long time~\cite{StrainCal,JusticePcal,EvanPcal,VirgoPcal}. The work reported here expands upon previous efforts ~\cite{GlasgowPcal,GEOPcal,HildEffect} and addresses dominant systematic uncertainties that can arise from absolute laser power calibration, test mass angular displacement, localized elastic deformations induced by the photon calibrator laser beams, and temporal variations in interferometer signals used to sense displacements. We have demonstrated methods devised to reduce or eliminate these major, and other smaller, uncertainties, reducing the overall voice coil actuator calibration uncertainty to less than 2 percent ($1\sigma$).

A key advantage of photon calibrators is that they can be used while an interferometer is running in its most sensitive operational state, referred to within LIGO as the {\it science-mode} configuration. Measuring in this configuration eliminates systematic uncertainties that arise in other methods, such as the standard calibration techniques employed during previous LIGO science runs~\cite{OcalPaper}, which require different optical and electronic configurations. Photon calibrators can be used either to determine the differential-length sensitivity of the interferometer directly, or to measure the voice coil actuation functions. Characterization of the voice coil actuators is essential because they are a key component of the control loop sensitive to gravitational waves, further described below. An additional advantage of photon calibrators is their ability to measure the time delay in the length response of the interferometer.~\cite{PcalTiming}

The DARM servo loop uses the ETM voice coil actuators to null the error signal of the differential-length degree of freedom of the interferometer, $(L_x-L_y)-n\lambda$, where $L_x$ and $L_y$ are the lengths of the two Fabry-Perot arm cavities, $n$ is an integer, and $\lambda$ is the laser wavelength.~\cite{RicksRef} The interferometer's gravitational wave output signal is measured at the error point of the DARM servo loop. The detector operates in a closed-loop state, so the response of the DARM error signal is suppressed by the gain of the servo loop. Therefore, the effect of the closed-loop servo must be determined in order to reconstruct the magnitude of the differential motion detected by the interferometer. Reconstruction is accomplished by measuring transfer functions of the overall open loop gain and some of the components of the DARM servo loop, then calculating the response of the interferometer to differential motion of the arm cavities.

There are three main components of the DARM servo loop: interferometer differential-length sensing, a series of digital filters, and differential-length actuation via displacement of the ETMs by voice coil actuators. In practice, the interferometer photodetector analog signals are digitized, filtered digitally, and converted to analog signals to actuate the ETMs via voice coils. The actuation functions consist of the actuation electronics, the voice coil force actuators, and the force-to-length responses of the suspended optics.

The actuation function of the test mass voice coil actuation path has been measured using several methods. One class of methods relies directly on the wavelength of the laser light as a length reference in the calibration process.~\cite{OcalPaper} Another method uses frequency modulation of the laser light to determine the actuation function.~\cite{VCOpaper} The method that is the focus of this article uses the recoil of photons from an auxiliary laser to induce a calibrated modulation in the differential-length of the interferometer arms.

This article is organized as follows: in Section 2 we discuss the principles of photon calibration and some experimental considerations; in Section 3 we describe the experimental apparatus for the LIGO photon calibrators; Section 4 summarizes measurements and results with various photon calibrator configurations; Section 5 discusses uncertainty estimates, and Section 6 provides conclusions and the outlook for future applications.

\section{Principles of photon calibration}
\begin{figure}
	\begin{flushright}
		\includegraphics[width=0.70\textwidth]{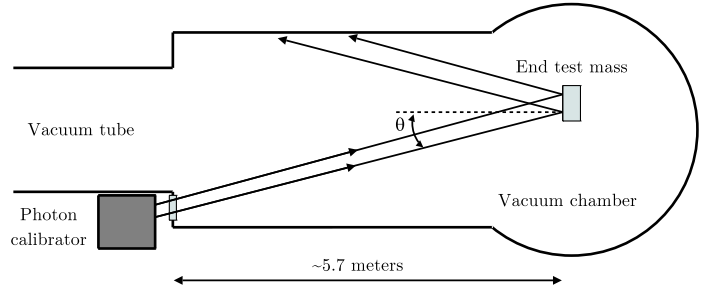}
	\end{flushright}
	\caption{Schematic diagram of a LIGO photon calibrator with output beams reflecting from an end test mass inside the vacuum envelope.}
	\label{fig:PcalAndVacuumChamber}
\end{figure}

When a beam of photons with time-dependent power $P(t)$ is incident upon the high reflectivity surface of an ETM at an angle of incidence $\theta$ (see Figure~\ref{fig:PcalAndVacuumChamber}), the beam reflects from the surface, transfers momentum from the recoiling photons, and thereby exerts a force on the mirror proportional to the power and the cosine of the angle of incidence. A sinusoidal power modulation can be written as,
\begin{equation}
P(t) = P_0 + P_m \sin(\omega t),
\label{eq:Power}
\end{equation}
where $P_0$ is the average power that is incident on the test mass, $P_m$ is the amplitude of the power modulation, and $\omega$ is the angular frequency.
The LIGO ETMs are suspended as pendulums with resonances at about 0.75 Hz. When the frequency of the modulated force on the optic is far above the pendulum resonance frequency, the optic is essentially free to move in the horizontal plane. For the photon calibrator, the amplitude of the induced motion, $x_m$, is given by
\begin{equation}
x_m(\omega) \simeq -\frac{2 P_m \cos\theta}{M c \omega^2},
\label{eq:Master_norotation}
\end{equation}
where $M$ is the mass of the mirror, $c$ is the speed of light, and the minus sign indicates the motion is 180 degrees out of phase with the applied force.

When the applied force is not directed through the center of mass of the optic, the induced torque causes an angular deflection of the test mass. The resonance frequencies for pitch and yaw rotations of the test mass are $\sim$0.5 Hz. Again, the mirror is freely rotating for modulation frequencies much greater than these resonance frequencies.

\begin{figure}
	\begin{flushright}
		\includegraphics[width=0.30\textwidth]{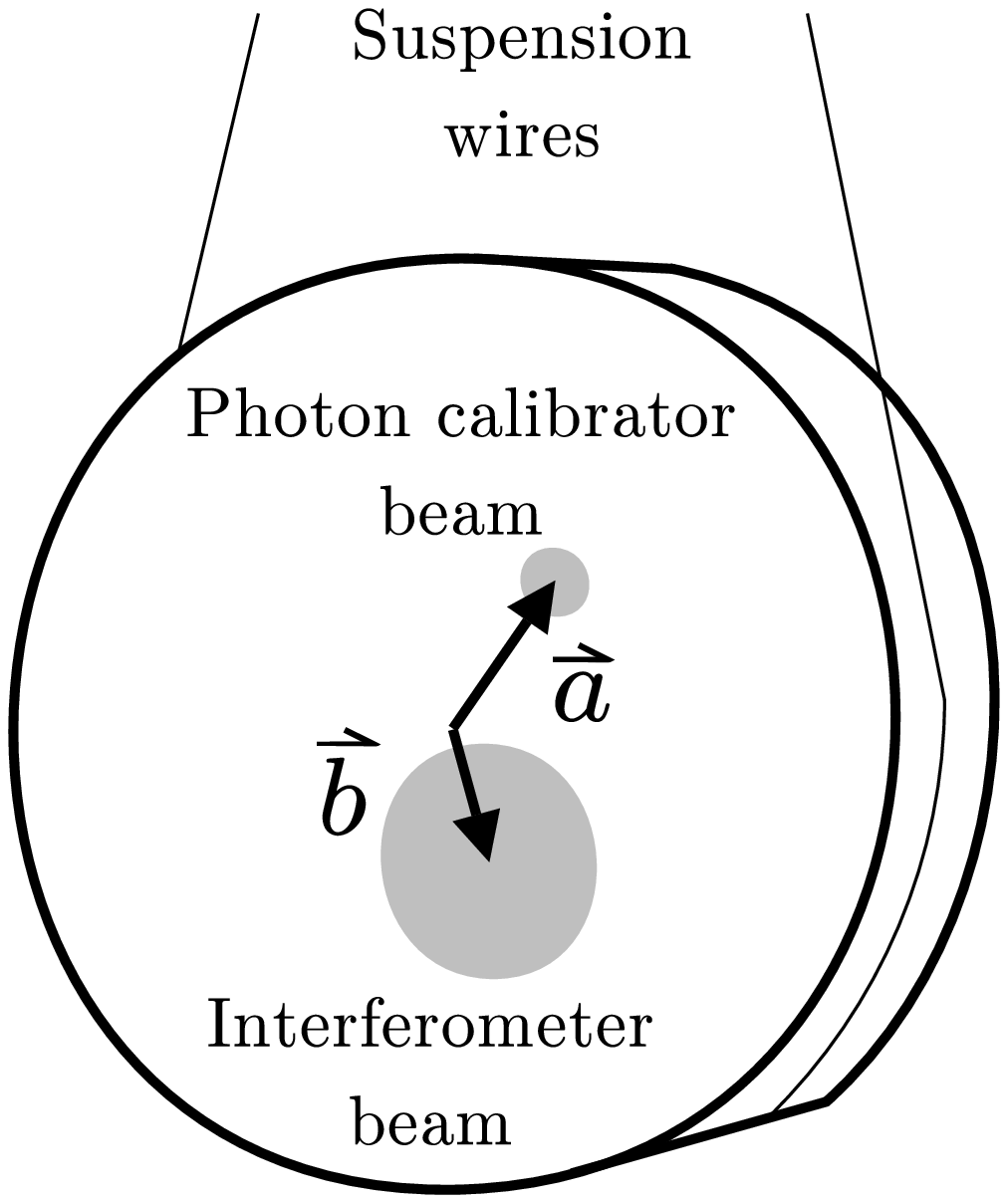}
		\includegraphics[width=0.60\textwidth]{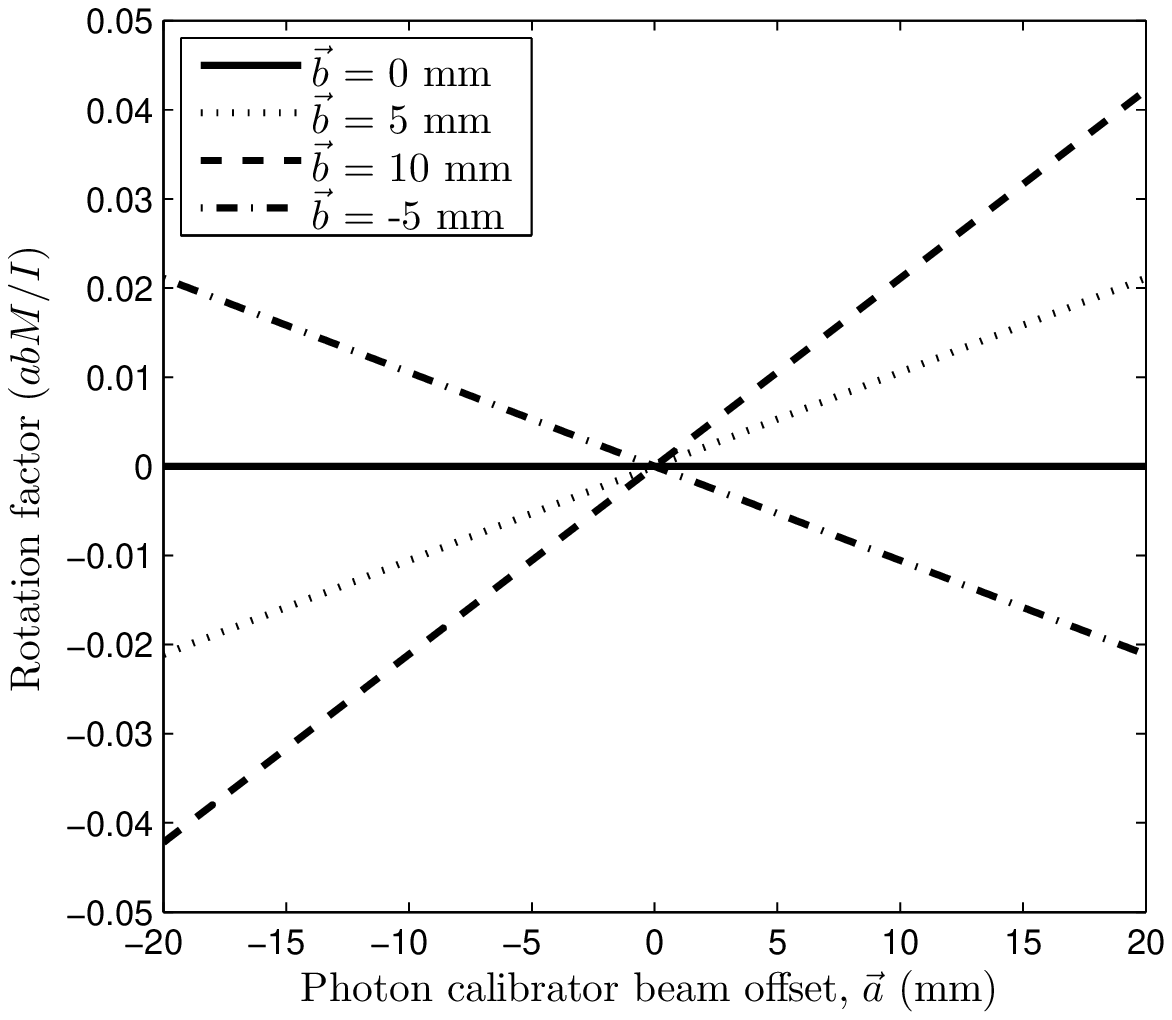}
	\end{flushright}
	\caption{Left: schematic diagram of a suspended ETM showing the locations of the photon calibrator and interferometer beams, with displacement vectors $\vec{a}$ and $\vec{b}$, respectively. The rotation-induced apparent length variations caused by the photon calibrator and sensed by the interferometer are proportional to $\vec{a}\cdot\vec{b}$. Right: rotation-induced length change versus photon calibrator beam offset for four interferometer beam displacements with $\vec{a}\,\|\,\vec{b}$.}
	\label{fig:PcalEtmRotation}
\end{figure}

Consider a photon calibrator beam that is incident at a point displaced from the center of the face of the optic, given by the displacement vector $\vec{a}$, as shown on the left in Figure~\ref{fig:PcalEtmRotation}. For small rotation angles, the induced torque is approximately $\vec{\tau} \simeq \vec{a} \times \vec{F}$, where $|\vec{\tau}| = aF$.

The equation of motion for the freely rotating optic is given by
\begin{equation}
I\ddot{\Omega}(\omega,t) = aF(\omega,t)
\end{equation}
where $\ddot{\Omega}$ is the angular acceleration. The optic is approximated as a right circular cylinder\footnote{The rear surface of the optic is actually wedged at 2 degrees. The maximum change in the rotational moment of inertia is approximately 0.2 percent; it is not included in this analysis.} with the moment of inertia about an axis through the center of the mass and parallel to the face of the optic given by $I=Mh^2/12+Mr^2/4$, where $M$ is the mass, $h$ is the thickness, and $r$ is the radius of the optic. For frequencies much greater than the rotational resonance frequencies, the modulated laser power induces a variation of the angle about the center of mass with amplitude given by
\begin{equation}
\Omega(\omega) \simeq -\frac{2 a P_m \cos\theta}{I c \omega^2}.
\end{equation}

If the interferometer beam is not centered, then the interferometer senses an apparent length change due to the rotation of the mass. For small angles of rotation, the effective length change, $x_{rot}$, is given by
\begin{equation}
x_{rot}(\omega) \simeq -\frac{2 \vec{a}\cdot\vec{b} P_m \cos\theta}{I c \omega^2}
\end{equation}
where $\vec{b}$ is the displacement vector of the center of the interferometer beam on the mirror's surface.

The effective length change due to the rotating mass adds or subtracts to the longitudinal length change, depending on the sign of $\vec{a}\cdot\vec{b}$. Thus, the total sensed motion due to the photon calibrator actuation is given by
\begin{equation}
x_{tot}(\omega) \simeq -\frac{2 P_m \cos\theta}{M c \omega^2} \left(1 + \frac{\vec{a} \cdot \vec{b} M}{I}\right).
\label{eq:Master}
\end{equation}
The factor $\vec{a}\cdot\vec{b}M/I$ for a LIGO ETM is plotted in the right-hand plot of Figure~\ref{fig:PcalEtmRotation} as a function of photon calibrator beam offset for various interferometer beam offsets. In this figure, the photon calibrator beam displacement is parallel to the interferometer beam displacement. For a photon calibrator beam offset by 10 mm in the same direction as an interferometer beam offset of 5 mm, the sensed length change due to rotation adds 1 percent to the total motion.

\begin{figure}
	\begin{flushright}
		\includegraphics[width=0.9\textwidth]{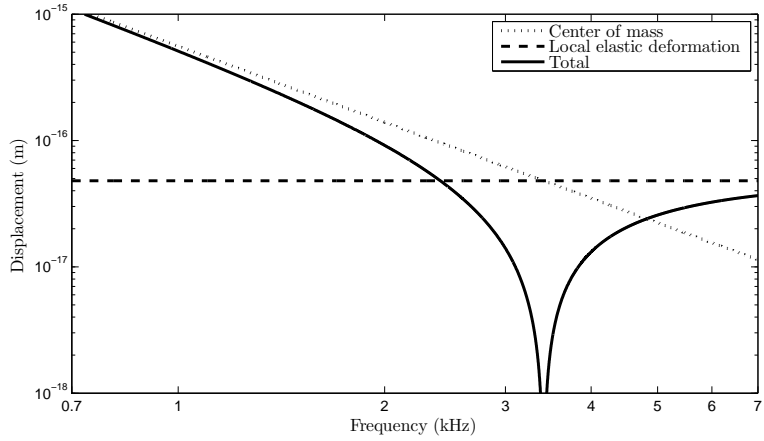}
	\end{flushright}
	\caption{The bulk displacement of the test mass as a function of frequency for a free mass (dotted line) falls as $f^{-2}$, the frequency-independent local elastic deformation (dashed line) 180 degrees out of phase with the bulk displacement, and the total surface motion (solid line) sensed by the interferometer. This is the same functional form as described by Hild, et al. The relative amplitude of these terms is taken from measurements in Section~\ref{sec:MeasAndResults}.}
	\label{fig:hildEffect}
\end{figure}

Hild et al. showed that localized elastic deformation of the test mass surface due photon calibrator radiation pressure can significantly change the amplitude of the sensed length modulation.~\cite{HildEffect} The free-mass motion falls as $f^{-2}$, but the elastic deformation is approximately frequency-independent for frequencies far below the test mass internal mode frequencies. The lowest internal mode frequency is approximately 6 kHz. At several kHz, the amplitudes of the free-mass motion and the elastic deformation are comparable (see Figure~\ref{fig:hildEffect}). The free-mass motion is 180 degrees out of phase with the force applied to the optic while the elastic deformation is in phase with the force applied. The amplitude of the sensed elastic deformation is strongly dependent on the overlap of the photon calibrator beam with the interferometer beam, and, to a lesser extent, the spatial intensity profiles of the beams and the specific shape and composition of the test mass. The interferometer is maximally sensitive to the elastic deformation caused by the photon calibrator when the interferometer and photon calibrator beam centroids are co-located on the face of the optic, while the effect is minimized when the beams do not overlap. Even at lower frequencies, the elastic deformation can contribute significantly to the sensed motion when the beams are closely located. If not properly accounted for, this introduces a frequency-dependent systematic error in the voice coil actuator calibration. For example, measurements performed using a single-beam photon calibrator (presented in Section~\ref{sec:MeasAndResults}) show that when the photon calibrator and interferometer beams are centered on the optic, the total sensed motion at 1 kHz is 10 percent smaller than the expected free-mass displacement, and indicate that near 3.4 kHz the elastic deformation of the surface is comparable to the free-mass motion (see Figure~\ref{fig:hildEffect}). Uncertainties in determining the beam positions can lead to significant errors in predicting the interferometer sensing of the elastic deformation.

To minimize the local elastic deformation effect, one can simply move the photon calibrator beam away from the region of the optic that the interferometer beam is sensing, typically the center of the optic. However, as previously discussed, the resulting torques would lead to undesired angular displacements. We instead use two beams, balanced in power and displaced symmetrically about the center of the face of the optic.

When using two laser beams, however, the ratio of powers of the beams becomes important. For two beams, the effective beam position can be described by 
\begin{equation}
\vec{a}_{eff} = \frac{\alpha\,\vec{a}_1 + \vec{a}_2}{\alpha+1}
\end{equation}
where the photon calibrator beam positions are $\vec{a}_1$ and $\vec{a}_2$ and the ratio of beam powers is $\alpha=P_1/P_2$. In practice, beam powers are adjusted such that $|1-\alpha|\leq0.02$ and the beams are positioned such that $\vec{a}_2=-\vec{a}_1$. Thus, $\vec{a}_{eff}=\vec{a}_1(\alpha-1)/(\alpha+1)$, which is typically less than $0.01\times\vec{a}_1$.

\section{Experimental setup}
\begin{figure}
	\begin{flushright}
		\includegraphics[width=0.66\textwidth]{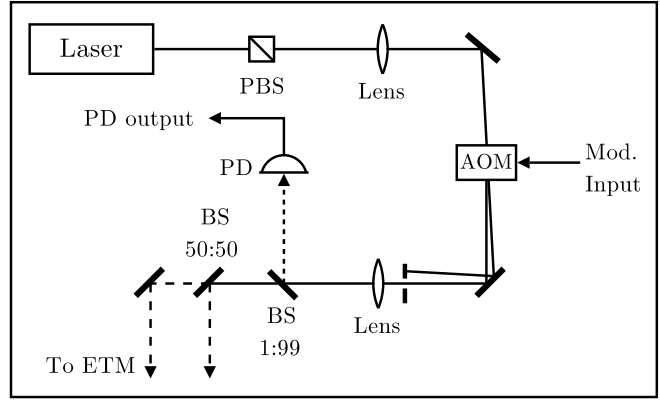}
	\end{flushright}
	\caption{Schematic diagram of the photon calibrator optical layout. PBS: polarizing beamsplitter, AOM: acousto-optic modulator, BS: beamsplitter, and PD: photodetector}
	\label{fig:PcalSchematic}
\end{figure}

Photon calibrators have been installed on each of the three LIGO interferometers, one to actuate each ETM. A schematic of a photon calibrator optical breadboard is shown in Figure~\ref{fig:PcalSchematic}. The horizontally polarized output of an optically-pumped $\textrm{Nd}^{3+}\textrm{:YLF}$ laser operating at a wavelength of 1047 nm is directed through a polarizing beamsplitter and is focused into an acousto-optic modulator that diffracts a fraction of the laser power that varies in response to the modulation input signal. The first-order diffracted beam is collimated and a sample is directed to a high-bandwidth, large-area germanium photodetector that provides a continuous monitor of the modulated laser power. The remaining beam is divided equally into two beams which are directed to the ETM.\footnote{The original layout for the LIGO photon calibrators was a single-beam configuration aligned to the center of the test mass to avoid large systematic errors due to rotation of the optic. They were converted to the two-beam configuration to avoid sensing the induced local elastic deformation.}

The optical breadboard is mounted on a platform outside of the vacuum enclosure. The photon calibrator beams enter the vacuum envelope through a glass viewport and impinge on the ETM as shown schematically in Figure~\ref{fig:PcalAndVacuumChamber}. The ETM is 25 cm in diameter, 10 cm thick, and has a mass of approximately 10.3 kg. The photon calibrator beams are displaced symmetrically by about 8 cm to either side of the center of the high reflectivity surface of the mirror. The interferometer spot size (radius) is about 3.4 cm. The angle of incidence of the two beams is approximately 9.6 degrees. The spot sizes (radii) of the two beams at the ETM surface are approximately 2 mm, and the average power of each beam is approximately 100 mW. The typical amplitude of the sinusoidally-modulated power in each beam is about 50 mW.

The positions of the photon calibrator beam spots are determined by observing the beams' scattered light on the ETM surfaces, using cameras mounted on other vacuum viewports. Accounting for parallax and refraction, alignment fiducials are provided by light emitting diodes (LEDs) used for mirror positioning, which emit light from apertures located close to the back surfaces of the optics. These fiducials are also used to determine the position of the interferometer spot on the face of the optic when operating in the science-mode configuration.

The photodetector that monitors the laser power is calibrated to indicate the laser power directed toward the vacuum window as a function of the voltage measured by the photodetector. To calibrate the photodetector, a power sensor that consists of an integrating sphere with a temperature controlled InGaAs photodetector and high-bandwidth current amplifier (working standard) is first calibrated against a second, identical power sensor (gold standard) which was sent to the National Institute of Standards and Technology (NIST) for absolute power calibration using 1047 nm $\textrm{Nd}^{3+}\textrm{:YLF}$ laser light~\cite{NISTcalibration}. Then, the working standard is used to measure the power exiting the photon calibrator. The optical efficiencies from incidence on the viewports to reflections from the ETMs were measured when the vacuum enclosures were open. For one ETM, for example, the overall optical efficiency is 90.7 percent. The viewport transmits 90.8 percent of the incident light and the ETM reflectivity is 99.9 percent. The laser power reflecting from the ETM can be continuously monitored by computing the product of the photodetector signal with the overall optical efficiency coefficient. The uncertainty in the absolute power calibration is discussed in Section~\ref{sec:Errors}.

\section{Measurements and results}\label{sec:MeasAndResults}
To determine the voice coil actuation function, $A$, for an ETM, the photon calibrator and the voice coil actuators sinusoidally actuate the position of the optic while the interferometer is operating in the science-mode configuration. By driving both actuators simultaneously, systematic errors induced by time-varying interferometer parameters, such as optical gain, are minimized. The sine wave frequencies are separated by 0.1 Hz, close enough to minimize interferometer response function variations, but far enough apart to minimize either signal contaminating the other due to leakage in the amplitude spectral density (ASD) calculation.

Each actuation is detected by the interferometer as a length modulation, and the signal appears as a peak above noise in the ASD of the error signal of the DARM servo loop. During the measurement, the peak in the ASD of the photon calibrator photodetector output and the peak in the ASD of the digital excitation signal sent to the voice coil actuator are also measured. The transfer coefficient magnitude is calculated from each ratio of error signal peak to excitation channel peak. Dividing the two transfer coefficients relates the digital excitation of the voice coils to the photon calibrator photodetector signal. The ETM voice coil actuation function is calculated using the previously obtained calibration of the photodetector, the mass of the ETM, the angle of incidence of the photon calibrator laser beams, the viewport transmission, the ETM reflectivity, the frequency of modulation, and the positions of the photon calibrator and interferometer laser beams on the ETM surface. The separation in frequency of the actuations requires a small correction for the frequency-dependent responses of the interferometer and the force-to-length actuation function.

\begin{figure}
	\begin{flushright}
		\includegraphics[width=0.75\textwidth]{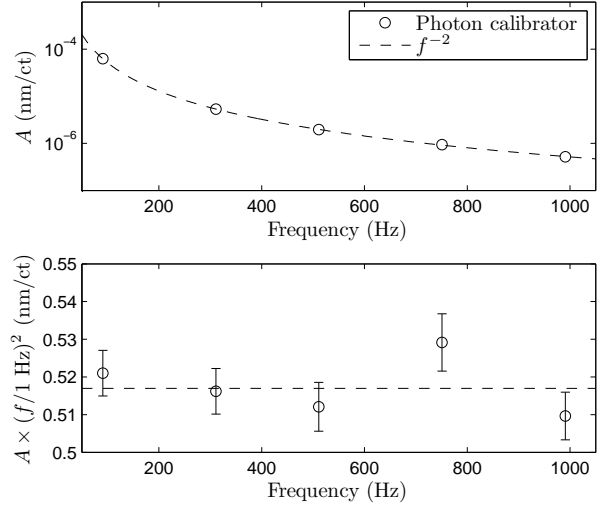}
	\end{flushright}
	\caption{H1 x-arm ETM voice coil actuation function measured with the photon calibrator versus frequency (upper panel). The dashed line indicates the expected $f^{-2}$ behavior. In the lower panel, the measured actuation function values are multiplied by the square of the measurement frequency. The error bars indicate the estimated $\pm1\sigma$ uncertainties.}
	\label{fig:pcalResults}
\end{figure}

The Hanford 4 km interferometer (H1) x-arm ETM voice coil actuation function has been measured at several frequencies between 90 Hz and 1 kHz, and the results are shown in the upper panel of Figure~\ref{fig:pcalResults}. At these frequencies, the ETM is essentially free to move in the longitudinal direction, so the force-to-length actuation function is expected to fall as $f^{-2}$. For comparison with this expected actuation function, the data are multiplied by the square of the measurement frequencies and plotted in the lower panel of Figure~\ref{fig:pcalResults} with their associated $\pm1\sigma$ error bars (see Section~\ref{sec:Errors}). A free-mass response would appear as a horizontal line in this plot. The peak-to-peak variation in these data is less than 3.7 percent.

\begin{figure}
	\begin{flushright}
		\includegraphics[width=1.0\textwidth]{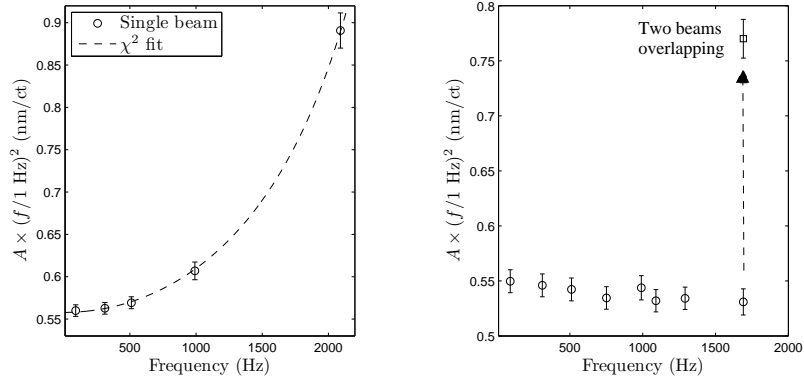}
	\end{flushright}
	\caption{Left panel: data from the single-beam, centered H1 y-arm photon calibrator with the $\chi^2$ fit to the data. Right panel: data from the two-beam H2 x-arm photon calibrator in two different configurations. First, with the beams are in their nominal positions, diametrically opposed about the center of the face of the optic, and, second, with the two beams overlapping with the interferometer beam at the center of the ETM surface. The error bars indicate the estimated $\pm1\sigma$ uncertainties.}
	\label{fig:hildResults}
\end{figure}

The influence of local elastic deformation for a single, centered photon calibrator beam is shown by the data in the left-hand plot of Figure~\ref{fig:hildResults}. Data from the single-beam H1 y-arm photon calibrator are shown for frequencies between 91 Hz and 2.1 kHz. The beam is centered on the test mass and overlaps the main interferometer beam. A chi-square fit to these data using the same functional form as Hild, et al.~\cite{HildEffect} is calculated from the data and their associated uncertainties. The model has two parameters, one for an idealized actuation function for a free mass falling as $f^{-2}$, and one for a frequency-independent deformation of the mirror surface that is 180 degrees out of phase with the free-mass motion. The calculated fit parameters are $5.58\pm0.04\times10^{-10}\,(1\,\textrm{Hz}/f)^2$ m/count for the free-mass response and $4.8\pm0.2\times10^{-17}$ m/count for the frequency-independent contribution of the surface elastic deformation to the sensed motion. The longitudinal displacement of the region of the optical surface sensed by the interferometer beam is thus less than the expected free-mass motion. If not accounted for, this results in a systematic error in the voice coil calibration that increases with frequency, as shown in Figure~\ref{fig:hildEffect}. For the H1 y-arm configuration, the discrepancy is approximately 60 percent at 2091 Hz (see Figure~\ref{fig:hildResults}).

Data for the Hanford 2 km interferometer (H2) x-arm photon calibrator using two beams symmetrically displaced from the center of the ETM are shown in the right-hand plot of Figure~\ref{fig:hildResults}. In this configuration, the expected $f^{-2}$ response (frequency-independent in this plot) is observed because the local elastic deformations caused by the photon calibrator beams are outside the region sensed by the interferometer beam. To confirm this, the two beams were aligned to the center of the optic, overlapping each other and the interferometer beam, and the voice coil actuation function was again measured at 1691 Hz. The actuation function increased by about 45 percent because the motion sensed by the interferometer is reduced due to the combined effect of the photon calibrator induced displacement and the elastic deformation of the optical surface. For the two beam configuration measurements on H2, the voice coil actuation function has a peak-to-peak variation of 3.6 percent about an idealized free-mass actuation function.

\section{Estimated uncertainties}\label{sec:Errors}
There are a number of potential sources of both statistical and systematic uncertainty that can impact the overall accuracy and precision of the voice coil actuation function derived from the measurements with the photon calibrators. We describe these sources below and summarize their respective estimated uncertainties in Table 1.
\begin{table}
	\caption{\label{tab:PcalErrors}Summary of the significant photon calibrator uncertainties for the H1 x-arm ETM voice coil calibration. Note that systematic errors arising from effects associated with test mass deformations have not been included.}
	\begin{indented}
	  \item[]\begin{tabular}{@{}ll}
	    \br
			Variable & $1\sigma$ uncertainty \\
			\mr
			Rotation, $(1+\vec{a}\cdot\vec{b}M/I)$ & 1.0 \% \\
			Power coefficient, $P_m$ & 0.7 \% \\
			Statistical ($N \simeq 100$) & 0.25 \% (Typical)\\
			Angle cosine, $\cos\theta$ & 0.1 \% \\
			Mass of ETM, $M$ & 0.1 \% \\
			\\
			\bf{Total error} & \bf{1.3 \%} \\
			\br
		\end{tabular}
	\end{indented}
\end{table}

A potentially large source of systematic uncertainty is the absolute power calibration of the photon calibrator's internal photodetector. The calibrated photodetector signal indicates the laser power directed toward the vacuum enclosure window. The estimated overall uncertainty in its calibration is 0.56 percent including contributions from several sources added in quadrature. The photodetector calibration relies on the absolute calibration, performed by NIST, of the gold standard power measurement system. The NIST calibration carries a $1\sigma$ uncertainty of 0.44 percent\footnote{Calibrations repeated yearly will indicate the long-term stability of the gold standard calibration.}.~\cite{NISTcalibration} To assess the variability in transferring the NIST calibration of the gold standard to the working standard, we repeated a detailed calibration procedure 25 times. The $1\sigma$ variation of the derived working standard calibration coefficients was 0.21 percent. The uncertainty introduced by variations in the positioning of the integrating sphere aperture relative to the incident laser beam was assessed by successive repositioning of the integrating sphere assembly. The $1\sigma$ variation of these measurements was 0.18 percent. Temporal variations in the calculated calibration of the internal photodetector were investigated by positioning the working standard in an installed photon calibrator output beam and simultaneously recording its output and the output of the internal photodetector over a two-week period. The $1\sigma$ variation of the ratio of the outputs, calculated via the 1-minute averaged output signals over two weeks was 0.22 percent. Adding these four contributions in quadrature gives the overall estimated uncertainty of 0.56 percent in the calibration of the internal photodetector in terms of the absolute power directed toward the vacuum viewport.~\cite{PcalError}

Measurements of the viewport transmission and ETM reflection coefficients were made with the working standard when the ETM vacuum enclosure was open. Measurements were made, both inside and outside the vacuum enclosure. These are relative measurements, so the working standard calibration uncertainty does not enter into these measurements. From these measurements, the estimated $1\sigma$ relative uncertainty in the overall optical efficiency is 0.40 percent. Combining in quadrature with the calibration of the photon calibrator photodetector gives a $1\sigma$ uncertainty of 0.69 percent in the absolute power reflecting from the ETM surface.

The position of the photon calibrator beams on the ETM surface can also be a major source of uncertainty when calculating rotation-induced length changes. First, an image is recorded of the optic's surface showing the relative locations of the photon calibrator beams and the LEDs located behind the optic. Then, using the LED spots as fiducials, the positions of the photon calibrator beams are determined using image processing software, taking into account refraction and parallax. The measurements are made several times and an average value is calculated. With this technique, the positions of the photon calibrator beams on the ETM surface are determined to within $\pm5$ mm. The location of the larger interferometer beam is known to within $\pm10$ mm. The uncertainty due to the rotation-induced length change is calculated for positions mis-measured by 5 mm and 10 mm for the photon calibrator and interferometer beams, respectively, the worst-case scenario. Additionally, for photon calibrators with two beams, a power imbalance ratio of two percent has been incorporated into the uncertainty estimate because this variation would effect the torque applied to the mirror. In total, the resulting estimated $1\sigma$ value is typically 1.0 percent. The uncertainty scales with the nominal locations of the photon calibrator and interferometer laser beams on the ETM surface.

The ETM mass was first calculated from the dimensions of the ETM and the density of the ETM substrate material. The masses of four ETMs (the x- and y-arms of H1 and the Livingston 4 km interferometer, L1) were measured using calibrated scales. The maximum discrepancy between the measured and calculated values is less than 20 grams. We use a rectangular window of 0.2 percent for the ETM mass that results in an estimated $1\sigma$ uncertainty of about 0.1 percent.

There are well constrained limits on the angle of incidence due to the physical constraints of the LIGO vacuum chamber assembly. The location of each photon calibrator beam spot on the viewport window is known to within 6 mm. The angle of the beams from the viewport spot location to the ETM spot location is determined from as-built technical drawings of the vacuum enclosure and the ETM placement. Calculations allow for a variation of 6 cm in the position of the ETM along the beam tube axis and 3 cm perpendicular to the beam tube axis, relative to the as-built technical drawings. Under these considerations, the cosine of the angle of incidence has a $1\sigma$ uncertainty of 0.1 percent.

Due to fluctuations of the optical gain of the interferometer, the interferometer differential-length sensitivity varies slightly as a function of time. These fluctuations induce a $1\sigma$ uncertainty of only 0.01 percent in the voice coil actuator calibration because the peaks in the DARM servo error signal are measured simultaneously and are separated in frequency by only 0.1 Hz.

The statistical errors in the overall calibration has been estimated from the standard deviation of multiple measurements of the ratio of the peaks in the DARM servo loop error signal to the peaks in the excitation monitor signals. From these measurements, the standard error is estimated to be 0.25 percent. At low frequencies, fewer measurements were required to obtain a standard error at this level, while at higher frequencies more measurements are required. The statistical uncertainty is frequency dependent due to the fact that the motion induced by the photon calibrator falls as $f^{-2}$ and the differential-length sensitivity is decreasing as $f^{-1}$ above roughly 200 Hz. A typical number of measurements to obtain this level of precision is $N\simeq100$ between 90 Hz and 1 kHz for integration times of 128 seconds.

Combining estimates of both systematic and statistical uncertainties, the total estimated $1\sigma$ uncertainty in the calibration of the voice coil actuators is 1.3 percent, indicative of the accuracy achievable with the photon calibrator. With increased precision in the localization of the photon calibrator and interferometer beams on the ETM surface, the overall uncertainty could be reduced below 1 percent.

This uncertainty estimate does not include contributions that arise from elastic deformation of the test mass by the photon calibrator forces. Finite-element analysis of the motion of the optical surface in response to dynamic external forces is ongoing.~\cite{WillemsDeform,MahmudaDeform} Preliminary results suggest that at frequencies above a few kHz bulk deformation of the test mass significantly changes the motion sensed by the interferometer beam. As the excitation frequency increases, there is a dramatic increase in the discrepancy between the free-mass motion, which is falling as $f^{-2}$, and the deformation-induced motion, which is increasing as the internal mode resonance frequencies are approached. For the measurements presented in this article, errors due to bulk deformation appear to be less than 1 percent. We thus estimate that the total photon calibrator uncertainty, including errors due to bulk elastic deformation, could be as high as 2 percent at the highest frequencies. Careful modeling may enable correcting for bulk deformation caused by the photon calibrator, but significant uncertainties may remain due to uncertainties in determining beam positions.

\section{Conclusions and outlook}
We have implemented high-precision photon calibrators on the LIGO detectors and used them to measure the ETM voice coil actuation functions at frequencies from 90 Hz to 2.1 kHz.  Measurements made in both single- and two-beam configurations have confirmed the importance of the local elastic deformation of the mirror surface induced by the photon calibrator beams predicted by Hild, et al. The two-beam configuration has been shown to sufficiently minimize the calibration errors caused by this effect. We have considered mirror rotation induced by non-centered photon calibrator beams and derived expressions for the sensed longitudinal motion as a function of the product of the interferometer and photon calibrator beam offsets and the effective beam position if the power is unbalanced when using a two-beam photon calibrator.

Estimated measurement uncertainties have been reduced to approximately 1.3 percent ($1\sigma$) by incorporating several improvements. These include accounting for rotation-induced apparent length variations, accurate power measurement, utilizing a two-beam photon calibrator configuration, and exciting simultaneously at closely spaced frequencies. Other potential sources of systematic errors were reduced by careful measurement of the transmission of the vacuum windows, the reflectivity of the test masses, the angles of incidence on the mirrors, the positions of the interferometer and photon calibrator beams, and the masses of the ETMs. Statistical errors were reduced by multiple averages of power spectral densities calculated from long-duration time series. An in-depth comparison of the photon calibrator technique with the method traditionally employed by LIGO which relies on the wavelength of the laser light and with a method based on laser frequency modulation is ongoing.~\cite{Comparison}

Frequency-dependent variations in the actuation path electronics and test mass deformations induced by the voice coil forces can cause the actuation function to deviate from the expected $f^{-2}$ force-to-length response of a free mass. However, the data presented in Figure~\ref{fig:pcalResults} indicate that the peak-to-peak deviation is less than 3.7 percent over the frequency range from 90 Hz to 1 kHz.  Finite-element modeling of test mass deformations due to the applied photon calibrator forces should facilitate correction for deformation-induced systematic errors, significant at frequencies above a few kHz, enabling investigation of test mass actuation functions at even higher frequencies.

One of the key advantages of the photon calibrator is its ability to operate in the most sensitive science-mode configuration. It is capable of introducing calibrated differential length displacements using an actuator that is outside the closed DARM control loop and thus enabling both calibration of the ``in-loop'' actuators and direct calibration of the monitor point sensitive to gravitational waves, the DARM servo error point. This capability, together with their demonstrated levels of precision and accuracy, makes photon calibrators a prime candidate for calibration of future gravitational wave detectors that will utilize more sophisticated test mass suspensions with more complex actuation chains.

\ack We gratefully acknowledge the LIGO calibration team and M. Rakhmanov for insightful discussions. We thank A. Effler, C. Gray, D. Hoak, and D. Lormand for their technical assistance and J. Hadler at the National Institute of Standards and Technology for his advice and assistance with absolute power calibration. We gratefully acknowledge the support of the National Science Foundation under grant PHY-0555406. LIGO was constructed by the California Institute of Technology and Massachusetts Institute of Technology with funding from the National Science Foundation and operates under cooperative agreement PHY-0107417. This document has been assigned LIGO Laboratory document number P080118.

\end{document}